\documentclass[submission, Proceedings]{SciPost}

\hypersetup{
    colorlinks,
    linkcolor={red!50!black},
    citecolor={blue!50!black},
    urlcolor={blue!80!black}
}

\usepackage[bitstream-charter]{mathdesign}
\DeclareSymbolFont{usualmathcal}{OMS}{cmsy}{m}{n}
\DeclareSymbolFontAlphabet{\mathcal}{usualmathcal}

\newcommand{\llangle}{\Big\langle \!\! \Big\langle}
\newcommand{\rrangle}{\Big\rangle \!\! \Big\rangle}
\newcommand{\as}{\alpha_s}
\newcommand{\xx}{\underline{x}}

\begin{document}

\begin{center}{\Large \textbf{
Single-Logarithmic Corrections to Small-$x$ Helicity Evolution\\
}}\end{center}

\begin{center}
Y. Tawabutr\textsuperscript{1$\star$}
\end{center}

\begin{center}
{\bf 1} The Ohio State University, Columbus, OH, USA
\\
* tawabutr.1@osu.edu
\end{center}

\begin{center}
\today
\end{center}


\definecolor{palegray}{gray}{0.95}
\begin{center}
\colorbox{palegray}{
  \begin{tabular}{rr}
  \begin{minipage}{0.1\textwidth}
    \includegraphics[width=22mm]{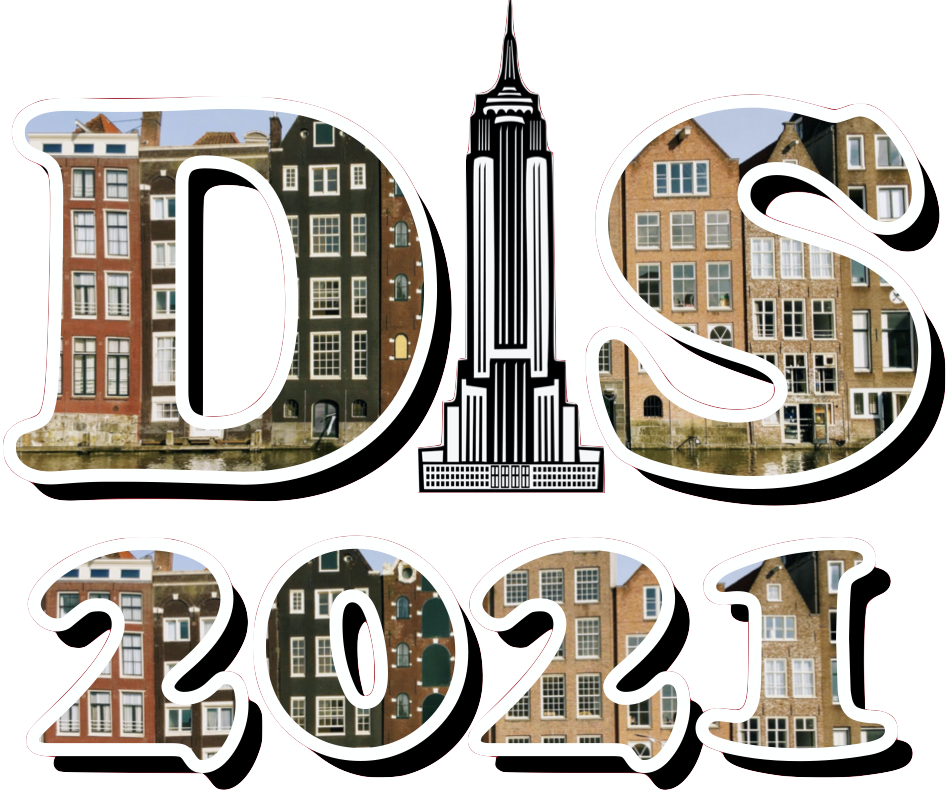}
  \end{minipage}
  &
  \begin{minipage}{0.75\textwidth}
    \begin{center}
    {\it Proceedings for the XXVIII International Workshop\\ on Deep-Inelastic Scattering and
Related Subjects,}\\
    {\it Stony Brook University, New York, USA, 12-16 April 2021} \\
    \doi{10.21468/SciPostPhysProc.?}\\
    \end{center}
  \end{minipage}
\end{tabular}
}
\end{center}

\section*{Abstract}
{\bf
The small-$x$ quark helicity evolution equations at double-logarithmic order, with the kernel $\sim\alpha_s\ln^2(1/x)$, have been derived previously. 
In this work, we derive the single-logarithmic corrections to the equations, to order $\alpha_s\ln(1/x)$ of the evolution kernel. The new equations include the effects of the running coupling and the unpolarized small-$x$ evolution, both of which are parametrically significant at single-logarithmic order. The large-$N_c$ and large-$N_c\& N_f$ approximations to the equation are computed. (Here, $N_c$ and $N_f$ are the numbers of quark colors and flavors, respectively.) Their solutions will provide more precise estimates of the quark helicity distribution at small $x$, contributing to the resolution of the proton spin puzzle.
}

\vspace{10pt}
\noindent\rule{\textwidth}{1pt}

\section{Introduction}
\label{sec:intro}

These proceedings are based on the work presented in \cite{LLA_2021} in more detail. 

Proton helicity receives contributions from spin and angular momenta of the constituent particles. For instance, the Jaffe-Manohar sum rule is \cite{Jaffe_Manohar}
\begin{align}
\frac{1}{2} &= S_q+S_G+L_q+L_G
\label{intro1}
\end{align}
where $S_q$ ($S_G$) is the spin angular momentum of the quarks (gluons) and $L_q$ ($L_G$) is the orbital angular momentum of the quarks (gluons). In this work, we mainly focus on the contribution coming from the spin of the quarks, i.e. the first term of Eq. \eqref{intro1}. At a given virtuality, $Q^2$, the quark's spin can be written as an integral over Bjorken-$x$ of the quark helicity distribution, $\Delta\Sigma(x,Q^2)$, which in turn can be expressed as a sum over the quark and antiquark flavors.
\begin{align}
S_q &= \frac{1}{2}\int_0^1dx\;\Delta\Sigma(x,Q^2) = \frac{1}{2}\int_0^1dx\sum_f\left[\Delta q_f(x,Q^2) + \Delta \bar{q}_f(x,Q^2)\right].
\label{intro2}
\end{align}
Experimentally, there are processes that allow us to extract $\Delta\Sigma(x,Q^2)$ in the range $x_{\min}\leq x\leq 1$ for some $x_{\min}>0$. Unless one can achieve an infinite-energy collision, it will not be possible to experimentally determine $\Delta\Sigma(x,Q^2)$ all the way down to $x=0$, in order to have a complete understanding of $S_q$ as required by the range of the integral in Eq. \eqref{intro2}. 

This project aims to fill in the gap by theoretically determining the asymptotic behavior of $\Delta\Sigma(x,Q^2)$ as $x\to 0$. In the small-$x$ dipole evolution framework, the asymptotic behavior can conveniently be determined through an evolution of a related quantity called ``polarized dipole amplitude.'' This object will be introduced in section \ref{sec:dip_amp}. Section \ref{sec:evol} presents the full evolution equations derived in term of the polarized dipole amplitude. The equations close in the large-$N_c$ and large-$N_c\& N_f$ limits as shown in section \ref{sec:closed}. Finally, the conclusion is given in section \ref{sec:conclusion}.

\section{Dipole Amplitudes}
\label{sec:dip_amp}

\begin{figure}[h]
\centering
\includegraphics[width=0.7\textwidth]{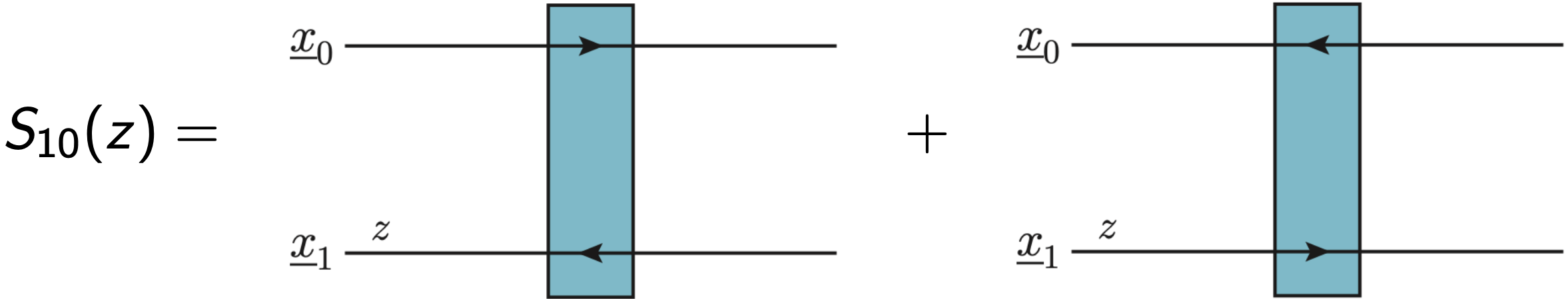}
\caption{Quark unpolarized dipole amplitude}
\label{fig:unpol_dip_amp}
\end{figure}

Consider a quark dipole with a quark (antiquark) at transverse position $\underline{x}_0$ and an antiquark (quark) at $\underline{x}_1$. Let $z$ be the longitudinal momentum fraction of the softer of the two (anti)quark lines. The dipole moves predominantly in the lightcone minus direction and eventually interacts with the target proton, which is moving predominantly in the lightcone plus direction. The amplitude of this interaction is called a quark (unpolarized) ``dipole amplitude,'' which can be written as
\begin{align}
S_{10}(z) &= \frac{1}{2N_c}\text{ Re}\left[\left\langle\mathcal{T}\text{ tr}\left[V_{\underline{0}}V_{\underline{1}}^{\dagger} \right]\right\rangle(z) + \left\langle\mathcal{T}\text{ tr}\left[V_{\underline{1}}V_{\underline{0}}^{\dagger} \right]\right\rangle(z) \right],
\label{dipamp1}
\end{align}
where $V_{\underline{n}}$ is the fundamental Wilson's line at transverse position $\underline{x}_n$ and the angle brackets denote averaging in the target's wave function. This quantity corresponds to the diagram in Fig \ref{fig:unpol_dip_amp}, where the blue rectangle represents the plus-moving target, i.e. the ``shockwave,'' whose lifetime is much shorter than that of the dipole. 

\begin{figure}[h]
\centering
\includegraphics[width=0.7\textwidth]{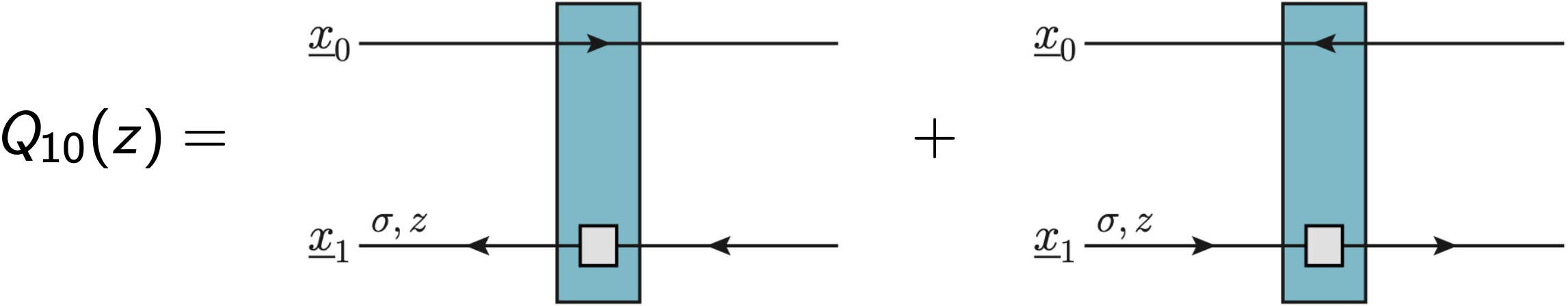}
\caption{Quark polarized dipole amplitude}
\label{fig:pol_dip_amp}
\end{figure}

In a similar setting of a minus-moving quark dipole interacting with a plus-moving target, one can consider the subleading part of the amplitude that depends on helicity, $\sigma$, of the (anti)quark at $\underline{x}_1$. This quantity defines the quark ``polarized dipole amplitude,'' which can be expressed as
\begin{align}
Q_{10}(z) &= \frac{zs}{2N_c}\text{ Re}\left[\left\langle\mathcal{T}\text{ tr}\left[V_{\underline{0}}V_{\underline{1}}^{\text{pol }\dagger} \right]\right\rangle(z) + \left\langle\mathcal{T}\text{ tr}\left[V_{\underline{1}}^{\text{pol}}V_{\underline{0}}^{\dagger} \right]\right\rangle(z) \right],
\label{dipamp2}
\end{align}
In Eq. \eqref{dipamp2}, $V_{\underline{1}}^{\text{pol}}$ is the fundamental ``polarized Wilson's line'' at $\underline{x}_1$, which is defined at the double-logarithmic order in \cite{Gluon_2017, QuarkOp_2018}, and $s$ is the center-of-mass energy squared for the projectile-target scattering. The factor of $zs$ in \eqref{dipamp2} follows from the fact that helicity-dependent interaction is sub-eikonal \cite{Quark_2016}. Diagrammatically, the quark polarized dipole amplitude is given in Fig \ref{fig:pol_dip_amp}, where the grey square marks the (anti)quark line whose helicity, $\sigma$, is tracked.

At small $x$ in the double-logarithmic approximation, the quark polarized dipole amplitude relates to the quark helicity distribution as follows \cite{Quark_num_2017}.
\begin{align}
\Delta\Sigma(x,Q^2) &= \frac{N_cN_f}{2\pi^3}\int^1_{\Lambda^2/s}\frac{dz}{z}\int_{1/zs}^{1/zQ^2}\frac{dx^2_{10}}{x^2_{10}}\int d^2\underline{b}\;Q_{10}(z),
\label{dipamp2a}
\end{align}
where $\underline{b} = \frac{1}{2}(\underline{x}_0+\underline{x}_1)$ and $\underline{x}_{10}=\underline{x}_1-\underline{x}_0$. Then, the small-$x$ asymptotic behavior of $\Delta\Sigma(x,Q^2)$ follows from the asymptotic solution to the evolution equations of $Q_{10}(z)$.

\begin{figure}[h]
\centering
\includegraphics[width=0.37\textwidth]{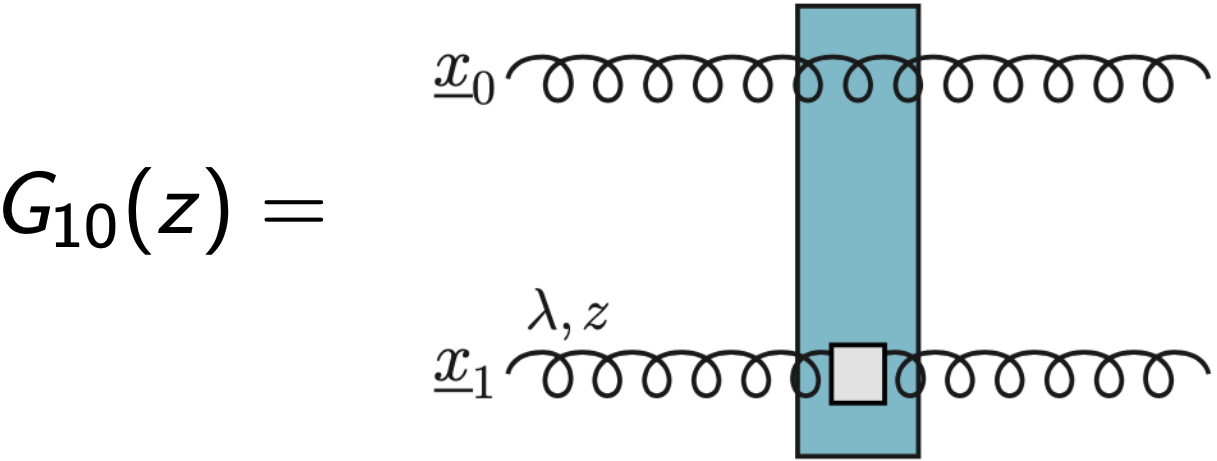}
\caption{Gluon polarized dipole amplitude}
\label{fig:glu_dip_amp}
\end{figure}

Finally, one can define the gluon polarized dipole amplitude in a similar fashion, starting from a gluon dipole with the dependence on one member's polarization, $\lambda$. The diagrammatic illustration is given in Fig \ref{fig:glu_dip_amp}, and the algebraic expression is
\begin{align}
G_{10}(z) &= \frac{zs}{2(N_c^2-1)}\text{ Re}\left[\left\langle\mathcal{T}\text{ tr}\left[U_{\underline{0}}U_{\underline{1}}^{\text{pol }\dagger} \right]\right\rangle(z) + \left\langle\mathcal{T}\text{ tr}\left[U_{\underline{1}}^{\text{pol}}U_{\underline{0}}^{\dagger} \right]\right\rangle(z) \right],
\label{dipamp3}
\end{align}
where $U_{\underline{n}}$ ($U_{\underline{n}}^{\text{pol}}$) is the adjoint (polarized) Wilson's line at transverse position $\underline{x}_n$. The gluon polarized dipole amplitude is important in our evolution equations because its evolution is mixed with that of the quark polarized dipole amplitude, as quarks emit gluons and vice versa. 

\section{Full Evolution Equations}
\label{sec:evol}

Consider quark and gluon polarized dipole amplitudes in Fig \ref{fig:pol_dip_amp} and \ref{fig:glu_dip_amp}. In these figures, the whole interaction between the dipole and the target is included in the shockwave. However, if we slightly relax the criteria of what is included in the shockwave, there will be an extra quark/gluon splitting occurring outside the target shockwave. If we treat such the splitting with lightcone perturbation theory (LCPT), $Q_{10}(z)$ and $G_{10}(z)$ can be shown to obey a system of integral equations with kernels of the form \cite{LLA_2021}
\begin{align}
\delta\left\langle\ldots\right\rangle &\sim \alpha_s\Bigg[\underbrace{\int\frac{dz'}{z'}\int\frac{dx^2_{21}}{x^2_{21}}}_{\text{DLA}} + \underbrace{\int\frac{dz'}{z'}\int dx^2_{21} }_{\text{SLA}_L} + \underbrace{\int dz'\int\frac{dx^2_{32}}{x^2_{32}}}_{\text{SLA}_T} + \ldots \Bigg]\otimes\left(\text{dipole amplitudes}\right),
\label{evol1}
\end{align}
where $z'$ and $\underline{x}_2$ are respectively the longitudinal momentum fraction and the transverse position of the daughter parton in the splitting. The three terms in the square brackets in Eq \eqref{evol1} correspond to different types of contributions to the kernel. In the first term, both transverse and longitudinal integrals are logarithmic, resulting in resummation of $\alpha_s\ln^2\frac{1}{x}$. We call this term ``double-logarithmic'' (DLA). In the second term, only the longitudinal integral is logarithmic, and we call the term ``single-logarithmic, longitudinal'' (SLA$_L$). Finally, in the third term, only the transverse integral is logarithmic, and we call the term ``single-logarithmic, transverse'' (SLA$_T$). The last two terms resum $\alpha_s\ln\frac{1}{x}$, which at small-$x$ is subleading to $\alpha_s\ln^2\frac{1}{x}$.

It is shown in \cite{LLA_2021, Quark_2016} that a splitting involving longitudinally soft parton emission results in a logarithmic longitudinal integral, which in turn leads to the explicit DLA and SLA$_L$ kernels. To derive the SLA$_T$ kernel, we have to consider hard-parton emission vertices. Putting everything together, the evolution equation for quark polarized dipole amplitude can be derived up to single-logarithmic order. For brevity, we only show one DLA + SLA$_L$ term and one SLA$_T$ term below in order to demonstrate important features of the equation.
\begin{align}
  & \frac{1}{N_c} \, \llangle \mbox{tr} \left[ V_{\underline{0}} \,
      V_{\underline{1}}^{\text{pol} \, \dagger} \right] \rrangle (z_{\min}, z_{\text{pol}}) =
  \frac{1}{N_c} \, \llangle \mbox{tr} \left[ V_{\underline{0}} \,
      V_{\underline{1}}^{\text{pol} \, \dagger} \right] \rrangle_0 (z_{\text{pol}}) \label{evol2} \\ 
      &+  \frac{1}{2 \pi^2} \int\limits_{\Lambda^2/s}^{z_{\min}} \frac{d z'}{z'}   \int\limits_{1/(z' s)} d^2 x_{2} \left[
    \frac{\as (1/x_{21}^2)}{x_{21}^2} \, \theta (x_{10}^2 z_{\min} - x_{21}^2 z') \,
    \frac{1}{N_c} \llangle \mbox{tr} \left[ t^b \,
        V_{\underline{0}} \, t^a \, V_{\underline{2}}^{\text{pol} \, \dagger} \right]
      \, U^{ba}_{\underline{1}} \rrangle (z', z') \right] \notag \\ 
    & \textcolor{blue}{ - \frac{1}{2 \pi^2} \int\limits_{0}^{z_{\text{pol}}} \frac{d z'}{z_{\text{pol}}} 
  \int\limits_{\frac{z_{\text{pol}}}{z'  (z_{\text{pol}} - z') s}} \frac{d^2 x_{32}}{x_{32}^2} \, \theta (x_{10}^2 z_{\min} z_{\text{pol}}  - x_{32}^2 z' (z_{\text{pol}} - z'))  \ \as \left( \frac{1}{x_{32}^2} \right) } \notag \\ & \textcolor{blue}{  \;\;\;\;\times  \, \left[ \frac{1}{N_c} \llangle \mbox{tr} \left[ t^b \,
      V_{\underline{0}} \, t^a \, V_{\underline{x}_1 - \frac{z'}{z_{\text{pol}}} \underline{x}_{32}}^{\dagger} \right]
    \, U^{\text{pol} \, ba}_{\underline{x}_1 + \left( 1 - \frac{z'}{z_{\text{pol}}} \right) \underline{x}_{32}} \rrangle ( \min \{ z_{\min}, z' , z_{\text{pol}}-z' \}  , z') \right]} \notag 
+ \left(\text{other terms}\right) ,
\end{align}
where the last two lines, colored in blue, correspond to SLA$_T$. The equation for gluon polarized dipole amplitude displays similar features, and it is written down in \cite{LLA_2021} together with the full quark equation. The two evolution equations are the main results of this work. In order to more clearly display the logarithmic structure of the kernel \cite{Quark_2016}, Eq \eqref{evol2} employs the double angle brackets notation, which are defined such that $\langle \! \langle\cdots\rangle \! \rangle = zs\langle\cdots\rangle$.

Another observation from Eq \eqref{evol2} is that the coupling constant, $\alpha_s$, is taken to be running. Since the QCD coupling constant runs with a single logarithm of the renormalization scale, $\mu$, its contribution potentially mixes with the helicity evolution kernel at the transverse single-logarithmic order in $x$. Hence, the running coupling contribution has to be identified and separated from the SLA$_T$ part of the kernel. The argument for the way $\alpha_s$ runs in each term requires more detailed consideration discussed in \cite{LLA_2021}.

Finally, an important feature of the evolution equation is that it is not closed. The objects on the right-hand side of the equations are often correlators of three Wilson lines and are, therefore, not dipoles, ultimately resulting in an infinite system of integral equations. With this issue at hand, solving the full equations directly for small-$x$ asymptotic of $\Delta\Sigma(x,Q^2)$ is difficult.

\section{Closed Evolution Equations}
\label{sec:closed}

Although the full evolution equations from section \ref{sec:evol} are not closed, the equations become closed once we take the large-$N_c$ or large-$N_c\& N_f$ limit. For instance, in the large-$N_c$ limit, consider the polarized dipole amplitude, integrated over the impact parameter $\underline{b} = \frac{1}{2}(\xx_0+\xx_1)$. Its evolution equation becomes
\begin{align}\label{eqn:evol3}
&G\left(x^2_{10},z_{\min},z_{\text{pol}}\right) = G^{(0)}\left(x^2_{10}, z_{\text{pol}}\right) + \frac{N_c}{\pi^2}\int\limits_{\Lambda^2/s}^{z_{\min}}\frac{dz'}{z'}\int\limits_{1/z's} d^2 x_2\\
&\;\;\;\;\times \left(\frac{\as(1/x^2_{21})}{x^2_{21}}\;\theta\left(x^2_{10}z_{\min}-x^2_{21}z'\right) - \as (\min\{1/x^2_{21},1/x^2_{20}\})\;\frac{\xx_{21}\cdot\xx_{20}}{x^2_{21}x^2_{20}}\;\theta\left(x^2_{10}z_{\min}-\max\left\{x^2_{21},x^2_{20}\right\}z'\right)\right) \notag \\
&\;\;\;\;\times \left[G\left(x^2_{21},z',z'\right) + \Gamma_{gen}\left(x^2_{20},x^2_{21},z',z'\right) \right]  \notag \\
&+ \frac{N_c}{2\pi^2}\int\limits_{\Lambda^2/s}^{z_{\min}}\frac{dz'}{z'} \int\limits_{1/z's}d^2 x_2 \, K_{\text{rcBK}} (\xx_{0}, \xx_{1}; \xx_{2}) \, \theta\left(x^2_{10}z_{\min}-x^2_{21}z'\right)    \left[G\left(x^2_{21},z',z_{\text{pol}}\right)  - \Gamma_{gen}(x^2_{10},x^2_{21},z',z_{\text{pol}}) \right]  \notag \\
&\color{blue} - \frac{N_c}{\pi^2}\int\limits_0^{z_{\text{pol}}}\frac{dz'}{z_{\text{pol}}}\int\limits_{1/z' s} \frac{d^2 x_{32}}{x_{32}^2} \;\as\left(\frac{1}{x^2_{32}}\right) \theta (x_{10}^2 z_{\min}   - x_{32}^2 z' ) \left[G\left(x^2_{32},z_{\min},z'\right) + \Gamma\left(x_{10}^2,x^2_{32},z_{\min},z'\right)  \right]  \notag \\
&\color{blue} + \frac{N_c}{2\pi^2}\int\limits_0^{z_{\text{pol}}}\frac{dz'}{z_{\text{pol}}} \left(2-\frac{z'}{z_{\text{pol}}}+\frac{z'^2}{z_{\text{pol}}^2}\right) \int\limits_{1/z' s} \frac{d^2 x_{32}}{x_{32}^2} \;\as\left(\frac{1}{x^2_{32}}\right) \theta (x_{10}^2 z_{\min}  - x_{32}^2 z' ) \,  \Gamma\left(x^2_{10},x^2_{32},z_{\min},z_{\text{pol}}\right)  , \notag
\end{align}
where $K_{rcBK}$ is the kernel for the unpolarized Balitsky-Kovchegov (BK) evolution equation with running coupling \cite{BK_Balitsky, BK_Kovchegov}. Here, $\Gamma(x^2_{10},x^2_{32},z_{\min},z_{\text{pol}})$ is called the neighbor dipole amplitude. It is an auxiliary function \cite{Quark_2016} whose large-$N_c$ evolution equation is 
\begin{align}\label{eqn:evol4}
&\Gamma\left(x^2_{10},x^2_{32},z_{\min},z_{\text{pol}}\right) = G^{(0)}\left(x^2_{10}, z_{\text{pol}}\right) + \frac{N_c}{\pi^2}\int\limits_{\Lambda^2/s}^{z_{\min}}\frac{dz'}{z'}\int\limits_{1/z's}d^2 x_4 \\
&\;\;\;\;\times \left(\frac{\as(1/x^2_{41})}{x^2_{41}}\;\theta\left(x^2_{32}z_{\min}-x^2_{41}z'\right) - \as (\min\{1/x^2_{41},1/x^2_{40}\})\;\frac{\xx_{41}\cdot\xx_{40}}{x^2_{41}x^2_{40}}\;\theta\left(x^2_{32}z_{\min}-\max\left\{x^2_{41},x^2_{40}\right\}z'\right)\right) \notag \\
&\;\;\;\;\times \left[G\left(x^2_{41},z',z'\right) + \Gamma_{gen}\left(x^2_{40},x^2_{41},z',z'\right) \right] \notag \\
&+ \frac{N_c}{2\pi^2}\int\limits_{\Lambda^2/s}^{z_{\min}}\frac{dz'}{z'}\int\limits_{1/z's} d^2 x_4 \, K_{\text{rcBK}} (\xx_{0}, \xx_{1}; \xx_{4}) \, \theta\left(x^2_{32}z_{\min}-x^2_{41}z'\right) \left[G\left(x^2_{41},z',z_{\text{pol}}\right)  - \Gamma_{gen}(x^2_{10},x^2_{41},z',z_{\text{pol}}) \right] \notag \\
&\color{blue} - \frac{N_c}{\pi^2}\int\limits_0^{z_{\text{pol}}}\frac{dz'}{z_{\text{pol}}}\int\limits_{1/z' s} \frac{d^2 x_{54}}{x_{54}^2} \;\as\left(\frac{1}{x^2_{54}}\right) \theta (x_{32}^2 z_{\min}  - x_{54}^2 z' ) \left[G\left(x^2_{54},z_{\min},z'\right) + \Gamma\left(x_{10}^2,x^2_{54},z_{\min},z'\right) \right] \notag \\
&\color{blue} + \frac{N_c}{2\pi^2}\int\limits_0^{z_{\text{pol}}}\frac{dz'}{z_{\text{pol}}} \left(2-\frac{z'}{z_{\text{pol}}}+\frac{z'^2}{z_{\text{pol}}^2}\right) \int\limits_{1/z' s} \frac{d^2 x_{54}}{x_{54}^2} \;\as\left(\frac{1}{x^2_{54}}\right) \theta (x_{32}^2 z_{\min}  - x_{54}^2 z') \,  \Gamma\left(x^2_{10},x^2_{54},z_{\min},z_{\text{pol}}\right)   . \notag
\end{align}
In Eq \eqref{eqn:evol3} and \eqref{eqn:evol4}, the last two lines of each equation, colored in blue, correspond to the SLA$_T$ terms. Furthermore, the generalized dipole amplitude, $\Gamma_{gen}(x^2_{10},x^2_{32},z_{\min},z_{\text{pol}})$, is defined as
\begin{align}
\Gamma_{gen}(x^2_{10},x^2_{32},z_{\min},z_{\text{pol}}) = G(x^2_{10},z_{\min},z_{\text{pol}})\theta(x_{32}-x_{10}) + \Gamma(x^2_{10},x^2_{32},z_{\min},z_{\text{pol}})\theta(x_{10}-x_{32}).
\label{evol5}
\end{align} 

With all these definitions, we now have a closed system of integral equations that can be solved to determine the high-energy asymptotic of $G(x^2_{10},z_{\min},z_{\text{pol}})$, which in turn implies the small-$x$ asymptotic of $\Delta\Sigma(x,Q^2)$ at large-$N_c$. Note that, at large $N_c$, the quark and gluon polarized dipole amplitudes satisfy $Q(x^2_{10},z_{\min},z_{\text{pol}}) = \frac{1}{4}G(x^2_{10},z_{\min},z_{\text{pol}})$. Hence, the equations no longer separate between quark and gluon dipoles in this limit.

In the large-$N_c\& N_f$ limit, the evolution equations reduce to a more complicated but closed system \cite{LLA_2021}. Despite the additional difficulty, the asymptotic behavior of the amplitudes can still be determined at least numerically.

\section{Conclusion and Outlook}
\label{sec:conclusion}

In this work, we summarize important features of the single-logarithmic contribution to the quark helicity evolution derived in \cite{LLA_2021}. The complete evolution equations are determined, but they are not closed. However, upon taking the large-$N_c$ or large-$N_c\& N_f$ limits, we obtain a closed system of equations we can solve, at least numerically, to determine the small-$x$ asymptotic of the quark helicity distribution. 

At double-logarithmic order (DLA), the closed equations have been solved both numerically \cite{Quark_num_2017} and analytically \cite{KPS_analytic_soln} for large-$N_c$ and only numerically for large-$N_c\& N_f$ \cite{KT_NcNf_soln}. In both regimes, the quark helicity distribution grows with $(1/x)^{\alpha_h}$ where $\alpha_h\approx 2.31$. However, in the large-$N_c\& N_f$ limit, the power term is multiplied by an oscillation factor corresponding to a periodic sign flip of the helicity distribution as $\ln(1/x)\to\infty$. Unfortunately, the period of the oscillation spans several units of rapidity and requires very high energy collisions to observe. Phenomenological implications of the large-$N_c$ solution are studied in \cite{pheno_2021}.

With the evolution equations known to the subleading single-logarithmic order, we are capable to determining the small-$x$ asymptotic of the helicity distribution more accurately. The amount of impact these additional SLA terms have on the small-$x$ asymptotic depends on phenomenological details and hence requires more thorough consideration. This is left for future work.


\section*{Acknowledgements}

The author would like to thank Yuri Kovchegov and Andrey Tarasov for collaborating on this project and Florian Cougoulic for discussions during the final stages of the work. Furthermore, the author would like to thank the DIS meeting organizers for the opportunity to present the work.

This material is based upon work supported by the U.S. Department of Energy, Office of Science, Office of Nuclear Physics under Award Number DE-SC0004286. The work is performed within the framework of the TMD Topical Collaboration.



\nolinenumbers

\end{document}